\documentclass[twocolumn,superscriptaddress,nobalancelastpage,amsmath,amssymb]{revtex4-1}
\usepackage{graphicx}
\graphicspath{{/}{Figures/}}
\usepackage{bm}
\usepackage{hyperref}

\usepackage{xcolor}

\newcommand {\nm}      {\,\mathrm{nm}}

\renewcommand {\vec}   {\mathbf}

\newcommand {\ket}[1]      {\lvert#1\rangle}

\newcommand {\avg}[1]   {\langle#1\rangle}

\newcommand {\half}       {\tfrac{1}{2}}
\newcommand {\threehalf}  {\tfrac{3}{2}}

\newcommand {\kdotp}   {\vec{k}\cdot\vec{p}}

\newcommand {\isopz}   {\tilde{\mathcal{P}}_z}
\newcommand {\isopy}   {\tilde{\mathcal{P}}_y}
\newcommand {\lB}      {l_B}
\DeclareMathOperator{\diag}{diag}

\begin{document}

\date{\today}

\title {Counterpropagating topological and quantum Hall edge channels}

\author{Saquib Shamim}
\altaffiliation{saquib.shamim@physik.uni-wuerzburg.de}
\affiliation{Experimentelle Physik III, Physikalisches Institut, Universit\"{a}t W\"{u}rzburg, Am Hubland, 97074 W\"{u}rzburg, Germany}
\affiliation{Institute for Topological Insulators, Universit\"{a}t W\"{u}rzburg, Am Hubland, 97074 W\"{u}rzburg, Germany}

\author{Pragya Shekhar}
\affiliation{Experimentelle Physik III, Physikalisches Institut, Universit\"{a}t W\"{u}rzburg, Am Hubland, 97074 W\"{u}rzburg, Germany}
\affiliation{Institute for Topological Insulators, Universit\"{a}t W\"{u}rzburg, Am Hubland, 97074 W\"{u}rzburg, Germany}

\author{Wouter Beugeling}
\affiliation{Experimentelle Physik III, Physikalisches Institut, Universit\"{a}t W\"{u}rzburg, Am Hubland, 97074 W\"{u}rzburg, Germany}
\affiliation{Institute for Topological Insulators, Universit\"{a}t W\"{u}rzburg, Am Hubland, 97074 W\"{u}rzburg, Germany}

\author{Jan B\"{o}ttcher}
\affiliation{Institut f\"ur Theoretische Physik und Astrophysik, Universit\"{a}t W\"{u}rzburg, Am Hubland, 97074 W\"{u}rzburg, Germany}

\author{Andreas Budewitz}
\affiliation{Experimentelle Physik III, Physikalisches Institut, Universit\"{a}t W\"{u}rzburg, Am Hubland, 97074 W\"{u}rzburg, Germany}
\affiliation{Institute for Topological Insulators, Universit\"{a}t W\"{u}rzburg, Am Hubland, 97074 W\"{u}rzburg, Germany}

\author{Julian-Benedikt Mayer}
\affiliation{Institut f\"ur Theoretische Physik und Astrophysik, Universit\"{a}t W\"{u}rzburg, Am Hubland, 97074 W\"{u}rzburg, Germany}

\author{Lukas Lunczer}
\affiliation{Experimentelle Physik III, Physikalisches Institut, Universit\"{a}t W\"{u}rzburg, Am Hubland, 97074 W\"{u}rzburg, Germany}
\affiliation{Institute for Topological Insulators, Universit\"{a}t W\"{u}rzburg, Am Hubland, 97074 W\"{u}rzburg, Germany}

\author{Ewelina M. Hankiewicz}
\affiliation{Institut f\"ur Theoretische Physik und Astrophysik, Universit\"{a}t W\"{u}rzburg, Am Hubland, 97074 W\"{u}rzburg, Germany}

\author{Hartmut Buhmann}
\affiliation{Experimentelle Physik III, Physikalisches Institut, Universit\"{a}t W\"{u}rzburg, Am Hubland, 97074 W\"{u}rzburg, Germany}
\affiliation{Institute for Topological Insulators, Universit\"{a}t W\"{u}rzburg, Am Hubland, 97074 W\"{u}rzburg, Germany}

\author{Laurens W. Molenkamp}
\altaffiliation{laurens.molenkamp@physik.uni-wuerzburg.de}
\affiliation{Experimentelle Physik III, Physikalisches Institut, Universit\"{a}t W\"{u}rzburg, Am Hubland, 97074 W\"{u}rzburg, Germany}
\affiliation{Institute for Topological Insulators, Universit\"{a}t W\"{u}rzburg, Am Hubland, 97074 W\"{u}rzburg, Germany}

\maketitle

\textbf{The survival of the quantum spin Hall edge channels in presence of an external magnetic field has been a subject of experimental and theoretical research.
The inversion of Landau levels that accommodates the quantum spin Hall effect is destroyed at a critical magnetic field, and a trivial insulating gap appears in the spectrum for stronger fields.
In this work, we report the absence of this transport gap in disordered two dimensional topological insulators in perpendicular magnetic fields of up to $16$\,T. Instead, we observe that a topological edge channel (from band inversion) coexists with a counterpropagating quantum Hall edge channel for magnetic fields at which the transition to the insulating regime is expected. For larger fields, we observe only the quantum Hall edge channel with transverse resistance close to $h/e^2$. By tuning the disorder using different fabrication processes, we find evidence that this unexpected $\nu=1$ plateau originates from extended quantum Hall edge channels along a continuous network of charge puddles at the edges of the device.}

\section*{Introduction}

Following the discovery of the quantum spin Hall effect in HgTe quantum  wells in 2007~\cite{KonigEA2007,BernevigEA2006}, the robustness of this effect in presence of an external magnetic field has been investigated theoretically and experimentally. Since an external magnetic field breaks the time reversal symmetry and the quantum spin Hall states are no longer topologically protected, it was believed that an arbitrarily small magnetic field would destroy the quantum spin Hall effect. Band structure calculations reveal that the inverted order of the lowest conduction-band and the highest valence-band Landau levels, characteristic of the quantum spin Hall phase, is maintained until a critical magnetic field $B^*$ is reached. At this field, the Landau levels cross and the quantum spin Hall phase transitions into a non-topological (also called as trivial) insulating state in the gap in the Landau level spectrum~\cite{TkachovHankiewicz2010,ScharfEA2012PRB,ChenEA2012PRB}. Experimentally, the investigation of this transition from topological to non-topological insulating phase can be complicated in narrow gap semiconductors like HgTe due to the formation of charge puddles \cite{VayrynenEA2013,VayrynenEA2014}. From microwave impedance microscopy measurements, Ma \emph{et al.} reported the observation of an unexpected edge conduction in topological HgTe quantum wells in the regime where one would expect an insulating state~\cite{MaEA2015NatComm}. The origin of this edge conduction was unclear.


\begin{table*}[tb]
\begin{tabular}{l|llllllll}
Device     & quantum & thickness       & Mn conc. & Dimensions           & Fabrication & $V_0$    & $R_{xx}^\mathrm{max}$    & $B^*$  \\
label      & well    & $d_\mathrm{QW}$ & $x$      & $L\times W$          & method      &          &                   &        \\
           &         &  (nm)           &          & ($\mu$m$\times\mu$m) &             & (V)      &  ($\Omega$)       & (T)    \\
\hline
Dev 1.d$1$ & QW1     &       $11$      & $2.4\%$  & $30\times 10$        & dry etch    & $-1.75$  & $2.7\times 10^6$  & $3.1$  \\
Dev 1.d$2$ & QW1     &                 &          & $600\times200$       & dry etch    & $-2.45$  & $3.0\times 10^6$  &        \\
Dev 2.d    & QW2     &       $11$      & $1.2\%$  & $30\times 10$        & dry etch    & $-2.2$   & $70\times 10^3$   & $6.7$  \\
Dev 2.w    & QW2     &                 &          & $30\times 10$        & wet etch    & $-0.48$  & $170\times 10^3$  &        \\
Dev 3.d    & QW3     &      $7.5$      & $0$      & $30\times 10$        & dry etch    & $-0.61$  & $47\times 10^3$   & $6.4$\\
\end{tabular}
\vspace*{5pt}
\caption{Devices presented in main text and extended data with device labels, quantum well number, thickness and Mn concentration, Hall bar dimensions, fabrication method (dry or wet etching), $V_0$ (voltage for maximal longitudinal resistance), $R_{xx}^\mathrm{max}$ (maximal longitudinal resistance at $B=0$) and $B^*$ (magnetic field above which the band inversion is lifted). The values are given at temperature $T=4.2$\,K for devices Dev 2.d and Dev 2.w, and at  $T=20$\,mK for all other devices.
}
\label{TableDevices}
\end{table*}


In this article, we demonstrate that fabrication induced disorder explains the experimentally observed absence of a gap in the Landau level spectrum of (Hg,Mn)Te and HgTe quantum wells with an inverted band structure. We use two distinct processes to fabricate the devices, leading to different levels of disorder: Firstly, a conventional dry-etching technique that previously was routinely used to fabricate HgTe microstructures~\cite{KonigEA2007,MaEA2015NatComm}, which is now known to degrade the mobility due to disorder-induced scattering \cite{BendiasEA2018}.
Secondly, a refined chemical wet-etching technique that has been specifically optimized to fabricate high-quality microstructures with suppressed scattering and consequently high mobility of charge carriers~\cite{BendiasEA2018}.
From low temperature magnetotransport measurements at different gate voltages (which tunes the chemical potential), we map out the Landau level spectrum for a large range of density of carriers and magnetic field of up to $16$\,T for the devices fabricated from the two methods. Remarkably, for the dry-etched device, instead of the insulating regime in the Landau level spectrum, which one would expect from band structure calculations, we observe a $\nu=1$ quantum Hall edge channel.
Additionally, we experimentally show that when the chemical potential is tuned into the bulk band gap, both topological (which originates from band inversion) and quantum Hall edge channels (from Landau levels) can coexist and counterpropagate at high magnetic fields ($5$--$10$\,T). While the topological edge channel is located at the physical edge of the device, the quantum Hall edge channel propagates along an equipotential line formed by the disordered network of charge puddles. The disorder being concentrated near the edge of the sample, it hosts a single quantum Hall edge channel, propagating in the opposite direction compared to the topological edge channel.
Magnetotransport measurements in wet-etched devices, which have much weaker disorder, do not show any evidence of counterpropagating edge channels and show the expected insulating regime in the Landau level spectrum, thus underlining the role of the charge puddles in the formation of quantum Hall edge channels. The coexistence of quantum Hall and the topological edge channels is related to the band inversion and is very different from counterpropagating charge transport in the fractional quantum Hall effect~\cite{LafontEA2019Science,MacDonald1990PRL,Moore_Haldane_1997PRB}.


\section{Results}

We have fabricated and measured multiple devices from (Hg,Mn)Te and HgTe quantum wells with varying layer thickness and composition. While the results reported here have been reproduced in nine devices, in this work, we present a representative subset of measurements on devices fabricated from the following quantum wells:
(1) Two (Hg,Mn)Te quantum wells which are $11$\,nm thick with a Mn concentration of $2.4$\% and $1.2$\% labelled as QW$1$ and QW$2$, respectively.
(2) A $7.5$\,nm thick HgTe quantum well labelled as QW$3$.
All three quantum wells have an inverted band structure. The topological phase space of (Hg,Mn)Te quantum wells identified by $\kdotp$ band structure calculations have been discussed in detail in Ref.~\cite{ShamimEA2020SciAdv}. QW$1$ is a direct gap topological insulator with a band gap $\sim 4.6$\,meV~\cite{ShamimEA2020SciAdv}. QW$2$ and QW$3$ have an indirect band gap of $\sim 11$ and $\sim 23$\,meV respectively (as determined from $\kdotp$ band structure calculations). While the low field transport properties are sensitive to the nature of the band gap (direct or indirect)~\cite{ShamimEA2020SciAdv}, the high field transport properties are nominally identical for both direct and indirect gap devices as will be shown here.
The devices fabricated from QW$1$ are labelled as Dev 1.d$1$ and Dev 1.d$2$; from QW$2$ are labelled as Dev 2.d and Dev 2.w; from QW$3$ are labelled as Dev 3.d, where d and w are used to indicate the dry- and wet-etching process respectively. The devices fabricated by dry etching use $110$\,nm of $\mathrm{SiO_2}$/$\mathrm{Si_3N_4}$ (grown by plasma enhanced chemical vapor deposition at $80\,{}^\circ$C) as gate dielectric while the wet-etched devices use $14$\,nm of $\mathrm{HfO_2}$ grown by atomic layer deposition at $30\,{}^\circ$C as the gate dielectric layer. In Table~\ref{TableDevices}, we list the devices with their relevant properties, in particular device dimensions and etching method. Unless otherwise mentioned, all magnetotransport measurements have been performed in a magnetic field $B$ perpendicular to the quantum well at a base temperature of $T = 20$\,mK.


\begin{figure}[tb!]
\includegraphics[width=1\linewidth]{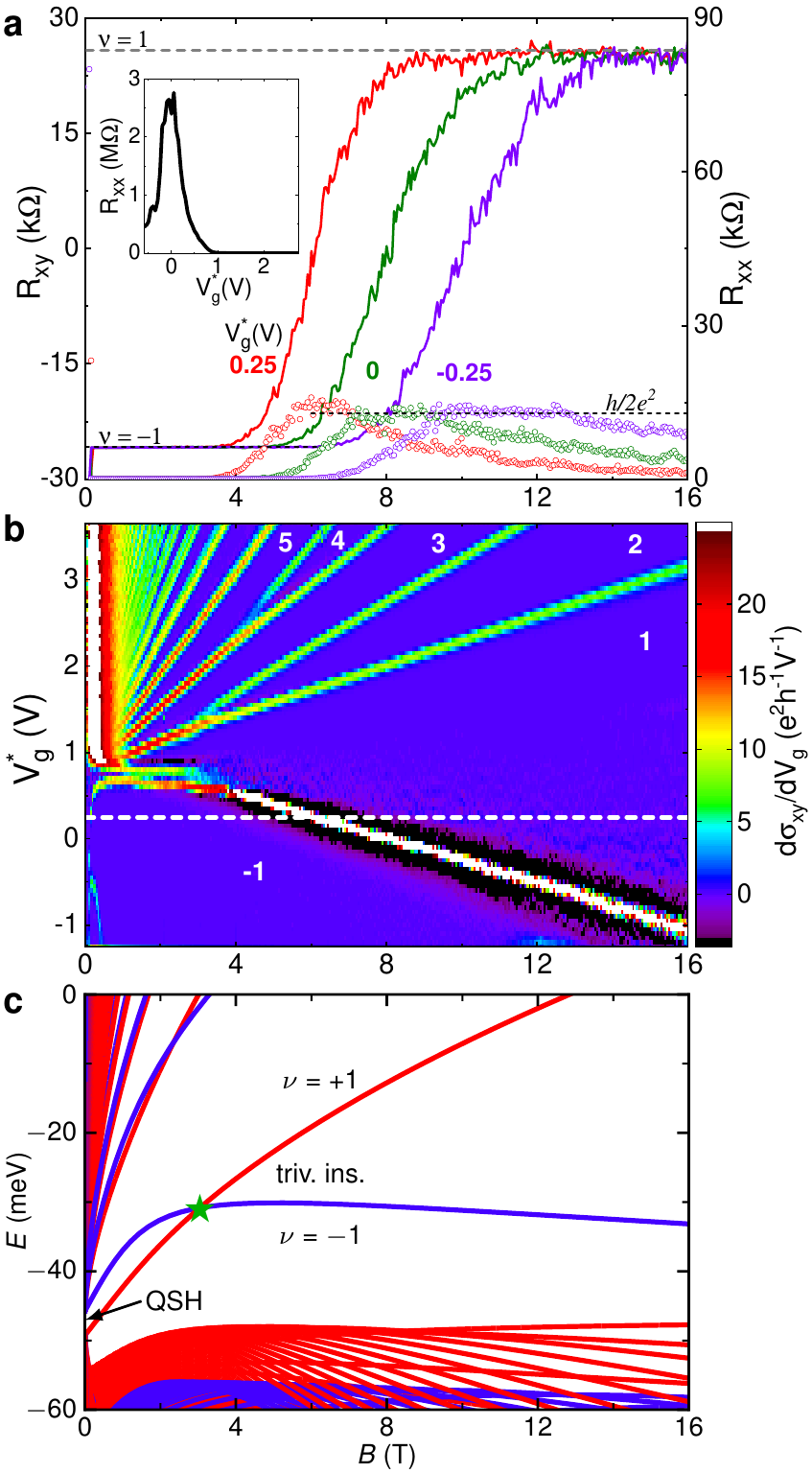}
\caption{\textbf{Absence of trivial insulating gap in Dev 1.d$1$.}
\textbf{a,} Transverse and longitudinal resistance, $R_{xy}$ and $R_{xx}$ respectively, as a function of perpendicular magnetic field $B$ at $20$\,mK for normalized gate voltages $V_g^*=0.25$, $0$, and $-0.25$\,V. The solid lines correspond to $R_{xy}$ (left axis) while the open circles correspond to $R_{xx}$ (right axis). The inset shows $R_{xx}$ as a function of $V_g^*$ at $20$\,mK.
\textbf{b,} Measured Landau level fan diagram showing $d\sigma_{xy}/dV_g^*$ as a function of $B$ and $V_g^*$ at $20$\,mK, where $\sigma_{xy}$ is the transverse conductivity. Numbers indicate quantum Hall filling factors $\nu$. The white dashed line shows a transition from $\nu=-1$ to $\nu=1$ quantum Hall plateau as a function of $B$ for a fixed $V_g^*$.
\textbf{c,} Calculated Landau level fan showing the Landau level crossing at $B=B^*$ indicated by a star. Red and blue colour distinguish the two blocks defined by isoparity (see Methods and Ref.~\cite{Beugeling2021PRB}).
`QSH' labels the quantum spin Hall effect at $B=0$\,T and `triv.\ ins.' labels the trivial insulator at large magnetic fields.
}
\label{FigMoto}
\end{figure}


Figure~\ref{FigMoto} shows the magnetotransport characteristics of Dev 1.d$1$ in a Hall bar geometry of length $L=30$\,$\mu$m and width $W=10$\,$\mu$m. The gate voltage characteristics (inset of Fig.~\ref{FigMoto}a) show that we can tune the chemical potential from $n$- to $p$-conduction regime (carrier type identified by classical Hall effect measurements) by decreasing the gate voltage $V_g$.
In order to compare among the devices which have different carrier densities at zero gate voltage, we have used the scaled gate voltage $V_g^*=V_g-V_0$, where $V_0$ is the gate voltage at maximal longitudinal resistance $R_{xx}$. In Table~\ref{TableDevices} we list $V_0$ for the devices presented in this work.
At $V_g^* = 0$\,V, $R_{xx}$ is larger than the expected value $h/2e^2$ for quantum spin Hall edge channels due to the relatively large dimensions of the device.
We have observed the quantized spin Hall resistance of $h/2e^2$ for microscopic devices fabricated using the wet-etching process from similar (Hg,Mn)Te quantum wells (see Supplementary Figure 1, Supplementary Note 1 and Ref.~\cite{ShamimEA2021NatComm}).
When the chemical potential is tuned in the bulk gap regime (around $V_g^*=0$\,V) or the valence band (negative $V_g^*$), the application of a small magnetic field ($> 50$\,mT) results in the formation of emergent quantum Hall plateaus, as evidenced by the quantization of the transverse resistance $R_{xy}$ to $-h/e^2$ (solid lines in Fig.~\ref{FigMoto}a) and simultaneous vanishing of the longitudinal resistance $R_{xx}$ (open circles in Fig.~\ref{FigMoto}a).
This low field behaviour is the same as reported in Ref.~\cite{ShamimEA2020SciAdv}, where we showed that these quantum Hall plateaus at very low magnetic fields emerge from the quantum spin Hall states enhanced by the band inversion-induced van Hove singularity in the valence band of topological (Hg,Mn)Te quantum wells. For $V_g^*=0$, the $\nu=-1$ plateau persists up to $B\sim 5$\,T. This value increases for more negative $V_g^*$, cf.\ Fig.~\ref{FigMoto}a.


\begin{figure}[t!]
\includegraphics[width=1\linewidth]{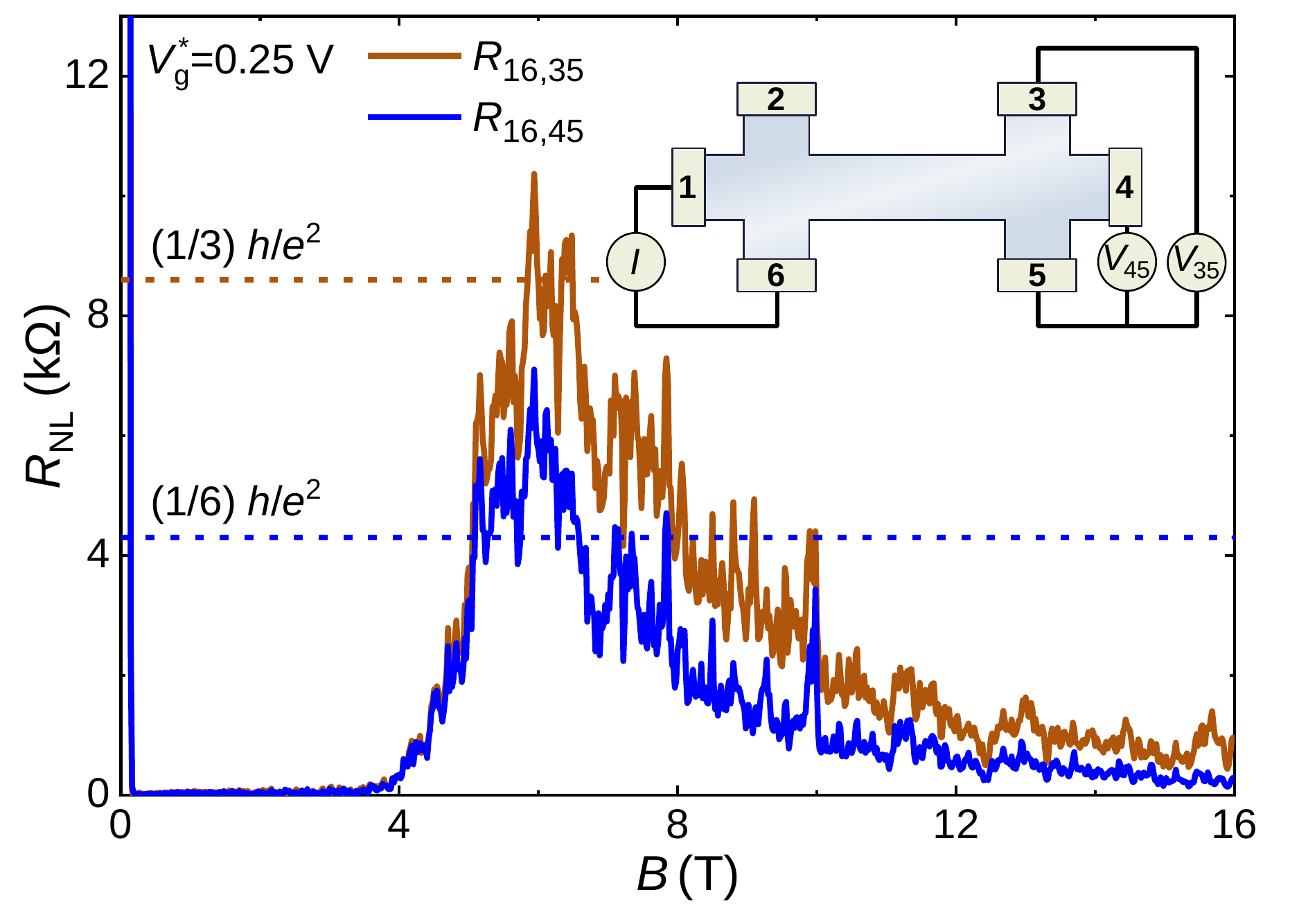}
\caption{\textbf{Nonlocal resistance measurement for Dev 1.d$1$.}
Nonlocal resistances  $R_{16,35(45)}=V_{35(45)}/I$ as functions of perpendicular magnetic field $B$ for $V_g^*=0.25$\,V. The inset shows the schematic of the measurement circuit where the current $I$ is applied across the contacts $1$ and $6$. The nonlocal voltages $V_{45}$ and $V_{35}$ are measured across contacts $4$ \& $5$ and $3$ \& $5$ respectively.
The dashed lines indicate the expected values for a counterpropagating pair of edge channels from the Landauer-B\"uttiker formalism.
}
\label{FigNonlocal}
\end{figure}


As $B$ increases further, $R_{xy}$ increases, changes sign and saturates to a $\nu=1$ quantized plateau characterized by $R_{xy}\approx h/e^2$ (dashed gray line in Fig.~\ref{FigMoto}a).
The sign of $R_{xy}$ and $\nu$ is normally associated with the charge carrier type (electrons or holes) in the system and hence a change in the sign as a function of $B$ (at constant $V_g^*$) is intriguing.
Interestingly, in the transition regime from $\nu=-1$ to $\nu=1$ plateau, $R_{xx}$ is quantized to $h/2e^2$ (dotted black line in Fig.~\ref{FigMoto}a) and decreases to a low value (few k$\Omega$ depending on $V_g^*$) at large $B$ when $R_{xy}$ is quantized to $h/e^2$. In addition, the $\nu=1$ plateau at high $B$ exhibits significant fluctuations (in $R_{xy}$) and $R_{xx}$ is still finite, in contrast to the emergent $\nu=-1$ and the regular quantum Hall plateaus of two-dimensional systems where $R_{xy}$ is essentially flat and $R_{xx}=0$ within the limits of experimental accuracy. The observed fluctuations in $R_{xy}$ at high $B$ are reproducible between measurements of the same device and result from quantum interference. The transition from $\nu=-1$ to $\nu=1$ is prominently visible in the Landau fan (Fig.~\ref{FigMoto}b), where we show $d\sigma_{xy}/dV_g^*$ as a function of $B$ and $V_g^*$. Similarly eye-catching is the complete absence of the insulating regime that we would expect from the calculated Landau fan shown in Fig.~\ref{FigMoto}c (calculation details in Methods section).


\begin{figure*}[tb]
\includegraphics[width=0.7\linewidth]{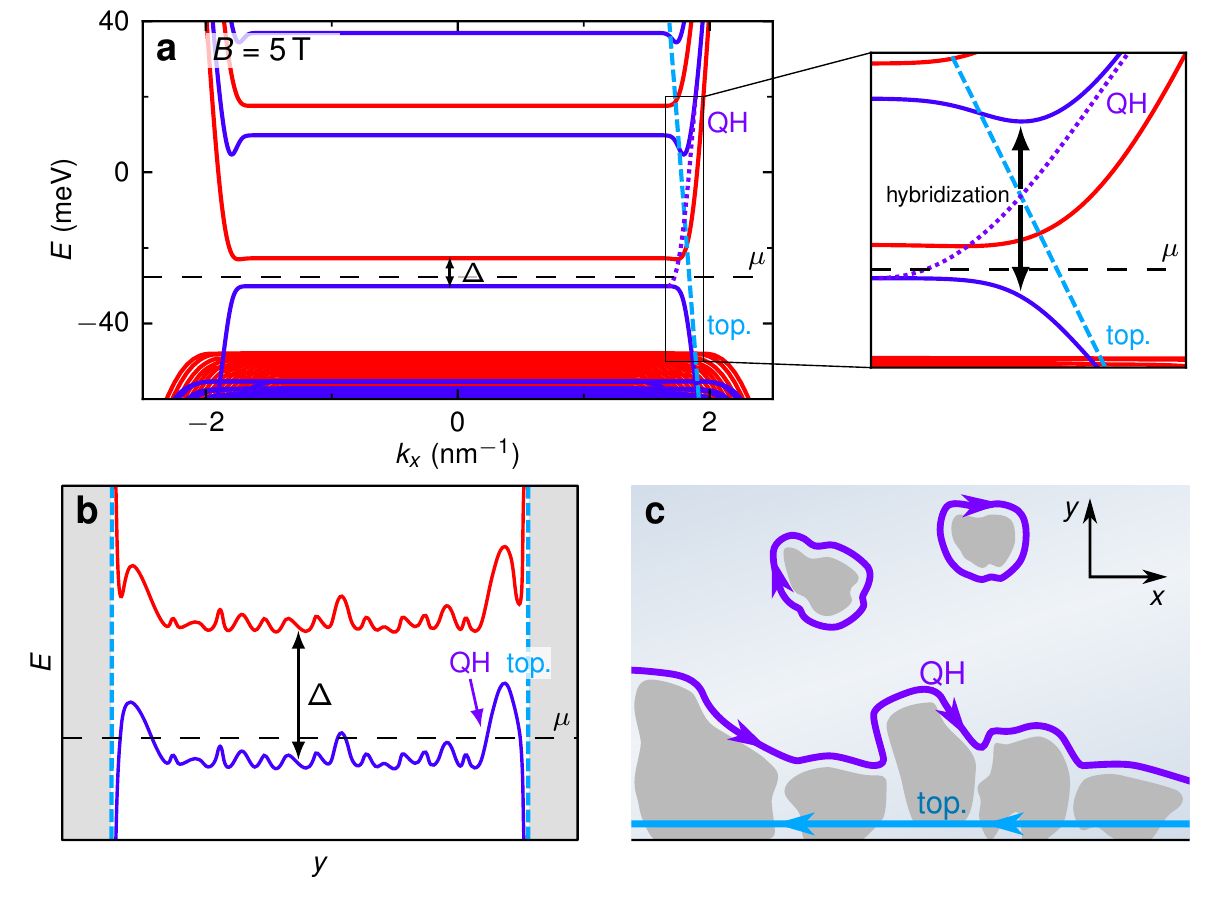}
\caption{\textbf{Formation of quantum Hall edge channels and the effect of charge puddles.}
\textbf{a,} Calculated band structure in strip configuration for a device similar to QW1, without disorder, in the trivial insulating regime at magnetic field $B=5$\,T. Red and blue colour distinguish the two blocks, cf.\ Fig.~\ref{FigMoto}\textbf{c} (see Methods and Ref.~\cite{Beugeling2021PRB}).
The dashed and dotted lines indicate the dispersions of the topological and quantum Hall edge channels in the topological block (blue) if hybridization were absent.
In the magnification, we show that strong hybridization in the topological block opens a gap between the topological (bright blue, dashed) and quantum Hall (violet, dotted) edge channel.
Thus, in absence of edge channels at the indicated chemical potential $\mu$, the system is trivially insulating.
\textbf{b,} Potential fluctuations (disorder) are added to the band structure of \textbf{a} along the width of the device ($y$ coordinate). Here $\langle y \rangle_{QH} = k_x l_B^2$ where $l_B=\sqrt{\hbar/eB}$ is the magnetic length. We emphasize the stronger fluctuations near the edges (vertical dashed lines).
\textbf{c,} Schematic showing a typical potential landscape at the chemical potential $\mu$ indicated in \textbf{b}, with a topological edge channel at the edge of the device (bright blue), the quantum Hall edge channel (violet) propagating along the network of the charge puddles, and localized channels in the bulk. We show only one edge of the device.
}
\label{FigSchematics}
\end{figure*}


The quantization of $R_{xx}$ at $h/2e^2$ is a clear indication of presence of counterpropagating edge channels~\cite{KonigEA2007}. Further convincing proof of the existence of counterpropagating edge channels can be obtained from nonlocal transport experiments~\cite{RothEA2009}. A schematic of such a nonlocal measurement setup in a six terminal Hall device is shown in the inset of Fig.~\ref{FigNonlocal}. A current $I$ flows through the contacts $1$ and $6$ while the nonlocal voltages $V_{45}$ and $V_{35}$ are measured across contacts $4$ \& $5$ and  $3$ \& $5$ respectively, which are spatially separated from the current path.
The nonlocal resistances $R_{16,35(45)}=V_{35(45)}/I$ as a function of $B$ for $V_g^*=0.25\,$V are shown in Fig.~\ref{FigNonlocal}. The nonlocal resistance is zero till $B \simeq 4$\,T, beyond which it increases, reaches the maximum at $B\sim 6$\,T and finally decreases, approaching zero at high $B$.
For a pair of counterpropagating edge channels, the Landauer-B\"{u}ttiker formalism~\cite{Buttiker1986PRL,RothEA2009} predicts $R_{16,35}=(1/3) h/e^2$ and $R_{16,45}= (1/6) h/e^2$, shown by the dashed lines in Fig.~\ref{FigNonlocal}. The agreement between the experimentally measured nonlocal resistance and the theoretically expected value confirms the existence of counterpropagating edge channels in the transition regime from $\nu=-1$ to $\nu=1$ plateau.
In contrast to the situation for the helical edge channels of the quantum spin Hall effect~\cite{BernevigZhang2006,BernevigEA2006} at zero field, there is no time-reversal symmetry between the edge channels at these high fields.
Here, the two edge channels have fundamentally different origins: the topological channel, associated to the $\nu=-1$ plateau, originates from the band inversion, while the quantum Hall channel, associated to the $\nu=1$ plateau, is generated by the magnetic field. Due to their distinct nature, their localization properties are different. The resulting spatial separation is key to the survival of the topological edge channel up to large magnetic fields and can be viewed as a manifestation of the parity anomaly in condensed matter physics~\cite{BoettcherEA2019,BoettcherEA2020}.


\begin{figure*}[tb!]
\includegraphics[width=1\linewidth]{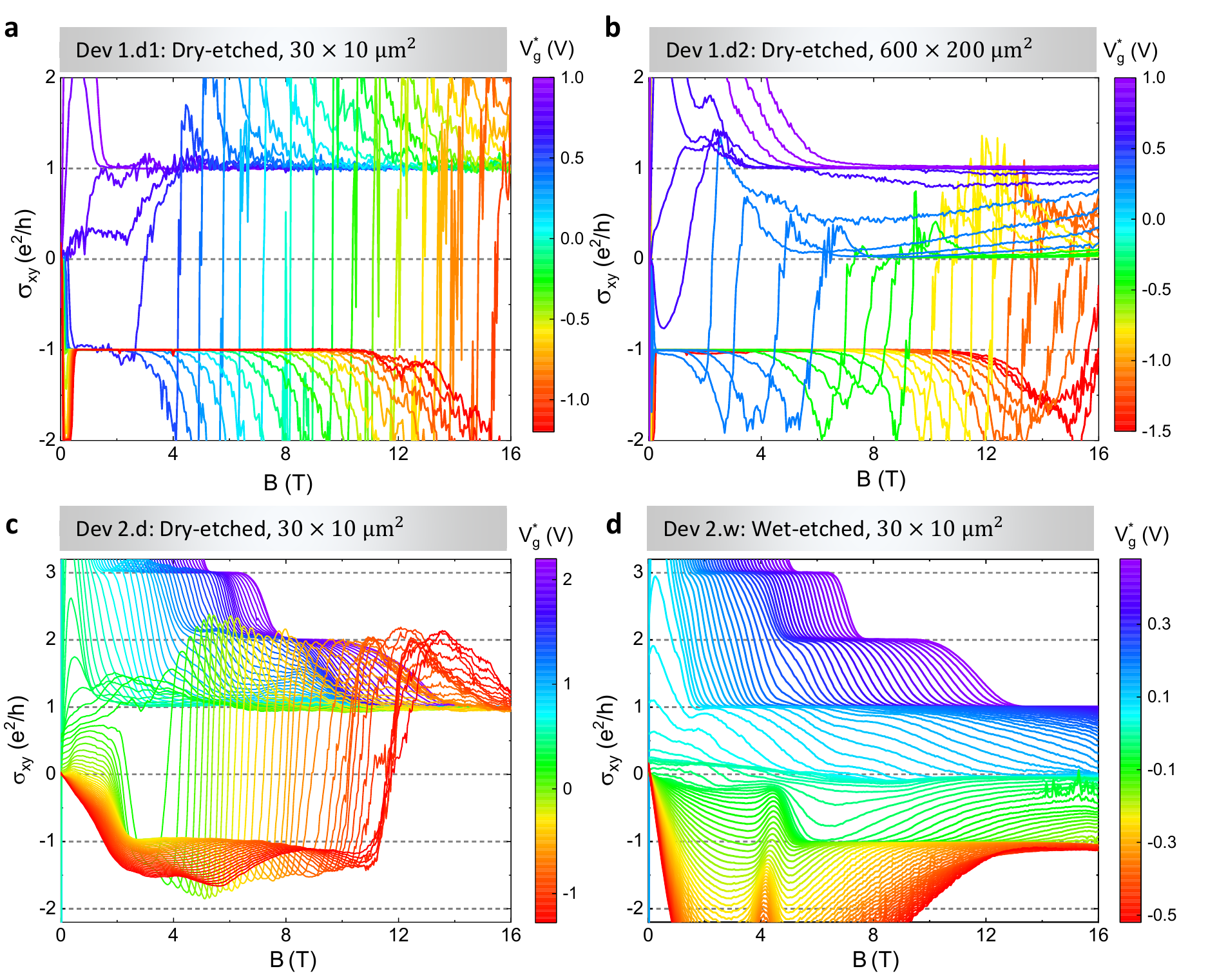}
\caption{\textbf{Comparison between devices of different dimensions and etching methods.}
The transverse conductivity $\sigma_{xy}$ as a function of perpendicular magnetic field $B$ for
\textbf{a,} Dry-etched device Dev 1.d$1$, size $30\times10\,\mu\mathrm{m}^2$ at $20$\,mK.
\textbf{b,} Dry-etched device Dev 1.d$2$, size $600\times200\,\mu\mathrm{m}^2$ at $20$\,mK.
\textbf{c,} Dry-etched device Dev 2.d, size $30\times10\,\mu\mathrm{m}^2$ at $4.2$\,K.
\textbf{d,} Wet-etched device Dev 2.w, size $30\times10\,\mu\mathrm{m}^2$ at $4.2$\,K.
The dashed lines show the integer values of $\sigma_{xy}$ in units of $e^2/h$.
}
\label{FigCompare}
\end{figure*}


A typical band structure calculated with the $\kdotp$ method at $B=5\,\mathrm{T}$ in a disorder-free sample is illustrated in Fig.~\ref{FigSchematics}a. The topological and quantum Hall edge channel, coming from the same block (blue in Fig.~\ref{FigSchematics}a), gap out as a result of hybridization. Thus, in absence of edge channels at the chemical potential $\mu$, a trivially insulating state would be expected. However, its absence in experiments indicates that disorder must not be neglected.

In narrow gap semiconductors, the defects at the interface of the semiconductor and insulator lead to local fluctuations in carrier density. When the gate voltage is used to tune the chemical potential, small charged regions form, known as ``charge puddles'' \cite{VayrynenEA2013,VayrynenEA2014}. It has been shown previously that such charge puddles are the dominant disorder for HgTe devices~\cite{LunczerEA2019PRL}.
Figure~\ref{FigSchematics}b shows a schematic of the potential fluctuations along the width of the device (cross section from one edge to the other edge).
The dry-etched devices (like Dev 1.d$1$) are patterned using lithography and subsequent physical etching using high energy ions. For such devices, the magnitude of the potential fluctuations is expected to be largest near the edges, where the etching takes place. Previous investigations of HgTe quantum wells (fabricated using a similar technology as in Ref.~\cite{MaEA2015NatComm}) showed that $p$-type charge puddles (regions in bulk that locally contain p-type charge carriers) are prevalent in these structures because of the large density of states near the top of the valence band~\cite{KonigEA2013PRX}. The formation of $p$-type puddles for a dry-etched device is schematically shown in Fig.~\ref{FigSchematics}b. In Fig.~\ref{FigSchematics}c we illustrate the high density of puddles near the physical edge of the device (gray areas). The edges of the dry-etched devices are formed by physically removing material using Ar ions. Hence the amount of disorder (density of puddles) is maximal near the edges. In strong magnetic fields, the quantum Hall edge channel (violet in Fig.~\ref{FigSchematics}c) propagates along an equipotential line of the disorder potential, facing the interior of the device, resulting in a propagation direction consistent with $\nu=1$. (Isolated charge puddles exist in the bulk of the material as well, that may host localized channels that do not contribute to transport at a macroscopic level.) In contrast, the exterior edge of the device hosts the topological edge channel (bright blue line), i.e., the single component of the quantum spin Hall edge channels that survives in strong magnetic fields, with opposite direction of propagation ($\nu=-1$). This topological edge channel is localized at the exterior edge regardless of disorder, because the localization is independent of momentum (see Methods). The spatial separation, a result of the distinct fundamental nature of the topological and quantum Hall edge channels in combination with disorder at the edge, suppresses scattering between them, and is thus a key ingredient for their coexistence.


\begin{figure}[!t]
\includegraphics[width=1\linewidth]{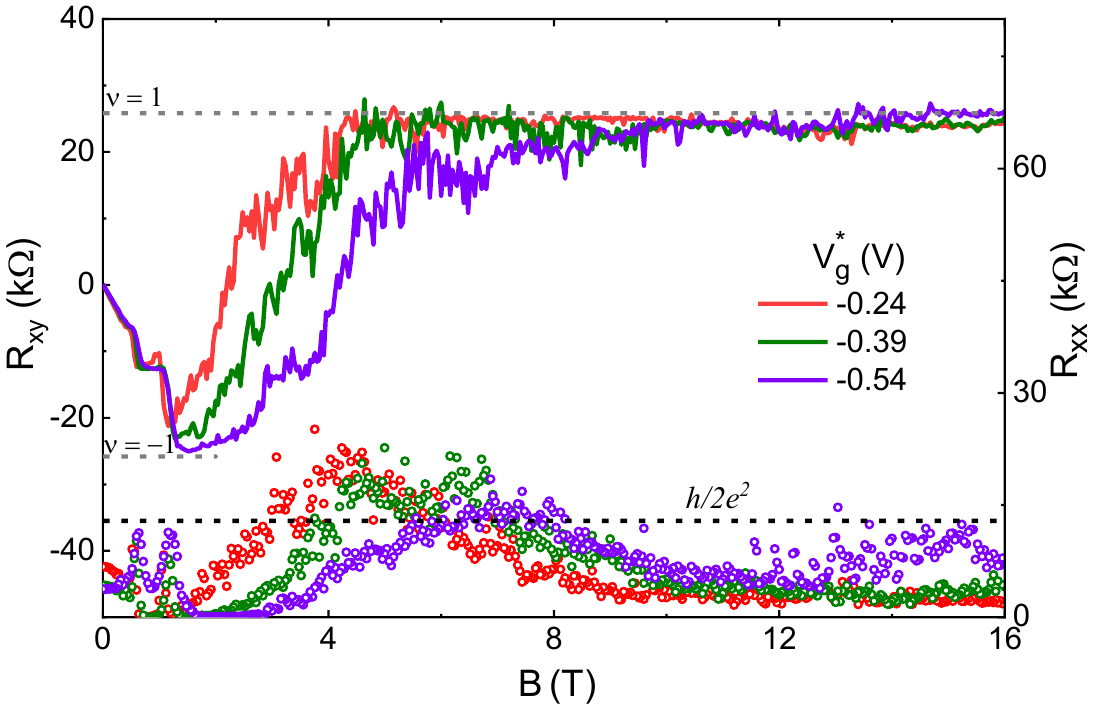}
\caption{\textbf{Counterpropagating edge channels and $\nu=-1$ to $\nu=1$ quantum Hall plateau transition in Dev 3.d.}
The transverse and longitudinal resistance, $R_{xy}$ and $R_{xx}$ respectively as a function of perpendicular magnetic field $B$ at $20$\,mK for gate voltages $V_g^*=-0.24$, $-0.39$, and $-0.54$\,V. The solid lines correspond to $R_{xy}$ (left axis) while the open circles correspond to $R_{xx}$ (right axis).
}
\label{FigMoto_MnFree}
\end{figure}


In order to confirm our hypothesis that charge puddles are essential for the observation of counterpropagating edge channels, we have performed two further experiments:
Firstly, since a continuous network of charge puddles along the entire edge of the device is essential to observe the quantum Hall edge channel, we perform magnetotransport measurements in devices of dimensions $600\times200\,\mu\mathrm{m}^2$ (Dev 1.d$2$) which are much larger than the dimensions $30\times10\,\mu\mathrm{m}^2$ of Dev 1.d$1$ presented in Fig.~\ref{FigMoto}.
At larger length scales, the possibility of formation of a continuous network of charge puddles is drastically reduced and we do not expect to observe the same signature of counterpropagating edge channels and the transition from $\nu=-1$ to $\nu=1$ quantum Hall plateau as a function of magnetic field \cite{BendiasEA2018}.
Indeed, we observe a weak signature of the transition from topological to quantum Hall plateau and an insulating state (region where $\sigma_{xy}=0$ at high magnetic fields) is distinctly visible in the $600\times 200\,\mathrm{\mu m^2}$ device (Fig.~\ref{FigCompare}b), unlike the small device (Fig.~\ref{FigCompare}a).
Secondly, we have compared devices fabricated using two different techniques: dry- and wet-etching. Since the wet-etching process is known to maintain the high quality of the material and the influence of charge puddles is expected to be significantly reduced as compared to dry-etched devices, we do not expect to observe the counterpropagating edge channels in the wet-etched devices (see calculated band structure and wave functions in Figs.~\ref{FigSchematics}a and b).
Figures~\ref{FigCompare}c and d show the transverse conductivity $\sigma_{xy}$ as a function of $B$ for dry- (Dev 2.d) and wet-etched (Dev 2.w) devices, respectively for various $V_g^*$. The dry-etched devices show a transition from $\nu=-1$ to $\nu=1$ plateau as a function of $B$ for $V_g^* \leq 0$\,V.
Importantly, the edge conduction is observed in the dry-etched devices in the entire range of the gate voltage and magnetic field investigated in the experiments, indicating the absence of a gap. In contrast, the wet-etched device shows that $\sigma_{xy}=0$ at high $B$ for various $V_g^*$ close to $0$\,V (Fig.~\ref{FigCompare}d). We identify this 
$\sigma_{xy}=0$ region as a non-topological insulating state, characterized by the absence of any edge channel, since the current through the device is zero and we do not observe any quantization of $R_{xx}$ to $h/2e^2$ (see Supplementary Figure 2).
From this we conclude that the counterpropagating edge channels are formed only in the dry-etched devices, while they are absent in the wet-etched ones. Thus, the amount of disorder (density of charge puddles) plays a crucial role in the formation of the extended quantum Hall edge channels.

The presence of Mn is not at all essential for the transition from $\nu=-1$ to $\nu=1$ quantum Hall plateau and the associated coexistence of topological and quantum Hall edge channels to occur. The HgTe and (Hg,Mn)Te quantum wells are identical in view of fundamental symmetry properties~\cite{Beugeling2021PRB} and Mn merely facilitates the observation of a $\nu=-1$ quantum Hall plateau at very low fields~\cite{ShamimEA2020SciAdv}. In Dev 3.d, a Mn-free HgTe two-dimensional topological insulator, we have observed the transition from $\nu=-1$ to $\nu=1$ quantum Hall plateau (Fig.~\ref{FigMoto_MnFree}). The previous observation of edge conduction in HgTe quantum wells at high magnetic field~\cite{MaEA2015NatComm} is likely due to the high density of charge puddles caused by lithographic and etching processes. Though a second chemical wet-etch process was done for devices in Ref.~\cite{MaEA2015NatComm} to obtain clean physical edges, the effect of the dry-etch process appears to be sufficient to give rise to edge channels at high magnetic fields.

\section{Conclusion}

To conclude, we have observed the absence of the theoretically predicted transport gap in the Landau level spectrum of disordered two-dimensional topological insulators at high magnetic field. We have shown the significant impact of the fabrication processes on the observed transport properties, in particular how the dry-etching process can lead to the observation of edge channels at high perpendicular magnetic fields where usually one would expect an insulating state. The observations confirm the crucial role of the charge puddles for the formation of a quantum Hall conductance channel, while also underlining the significance of the distinct fundamental nature of topological versus quantum Hall edge channels. Our findings further demonstrate that, in HgTe, sufficient lithographic control is now available to mitigate the influence of charge puddles on quantum transport. The simultaneous occurrence of counterpropagating topological and quantum Hall edge channels is a novel signature of materials with an inverted band structure.

\section{Methods}
\textbf{Material growth and characterization}:
We use commercial (Cd,Zn)Te substrates to grow HgTe and (Hg,Mn)Te quantum wells. Using molecular beam epitaxy, we grow a CdTe buffer layer of thickness $50$\,nm, followed by a (Hg,Cd)Te barrier, a HgTe or (Hg,Mn)Te quantum well, and another (Hg,Cd)Te barrier. X-ray diffraction measurements are used to determine the thickness of the quantum wells and the Mn concentration. The quantum wells show a typical mobility $\mu \sim 2\times 10^5\,\mathrm{cm}^2\,\mathrm{V}^{-1}\,\mathrm{s}^{-1}$ for carrier density $n \sim 5\times 10^{11}$\,$\mathrm{cm}^{-2}$.

\textbf{Device fabrication}:
The quantum wells are fabricated into Hall bars with different dimension using optical lithography and etching. We use two distinct etching techniques: conventional dry etching using high energy argon ions and chemical wet etching. The devices fabricated by dry etching use a $110$\,nm of $\mathrm{SiO_2}$/$\mathrm{Si_3N_4}$ (grown by plasma enhanced chemical vapor deposition at $80\,{}^\circ$C) as gate dielectric while the wet-etched devices use a $14$\,nm of $\mathrm{HfO_2}$ grown by atomic later deposition at $30\,{}^\circ$C as the gate dielectric layer. The gate electrode is formed by deposition of $5/200$\,nm of Ti/Au onto the dielectric. The ohmic contacts to the quantum well are realized by deposition of $50$\,nm AuGe and $50$\,nm Au by e-gun evaporation. Further details of the fabrication process can be found in Ref.~\cite{BendiasEA2018}.

\textbf{Magneto-transport measurements}:
The magnetotransport measurements have been performed in a $\mathrm{He_3}$-$\mathrm{He_4}$ dilution refrigerator of base temperature $20$\,mK. The longitudinal and transverse resistance, $R_{xx}$ and $R_{xy}$ respectively, have been measured using low frequency ($\sim 13$\,Hz) lock-in techniques in a four probe configuration. The current flowing through the device was in the range $2$--$20$\,nA.

\textbf{Band structure calculations with the $\kdotp$ model}:
We have used a $\kdotp$ model with a basis of eight orbitals \cite{Kane1957}:
$\ket{\Gamma_6,+\half}$, $\ket{\Gamma_6,-\half}$, $\ket{\Gamma_8,+\threehalf}$, $\ket{\Gamma_8,+\half}$, $\ket{\Gamma_8,-\half}$, $\ket{\Gamma_8,-\threehalf}$, $\ket{\Gamma_7,+\half}$, $\ket{\Gamma_7,-\half}$.
We have modelled the Hall bar geometry as a $500$\,nm wide strip with the layer stack as described in ``Material growth and characterization''. The system is spatially discretized in two dimensions, with the remaining (longitudinal in-plane) direction being represented as momentum $k$. The reduced width compared to the physical device does not affect the physics of the edge channels. The Hamiltonian matrices of dimension $\sim 10^6$ are diagonalized using sparse matrix algorithms available from the SciPy package for Python. The calculations were performed at the Julia-HPC cluster at the University of W\"urzburg.

The colour code in Figs.~\ref{FigMoto}c and \ref{FigSchematics}a indicates \emph{isoparity} $\isopz = P_z Q$, a combination of parity $P_z: z\mapsto-z$  in the perpendicular direction and a spin-like degree of freedom represented by the diagonal matrix $Q = \diag(+1,-1,+1,-1,+1,-1,-1,+1)$ in the orbital basis. This is a conserved quantity that divides the spectrum into two `blocks' defined by isoparity eigenvalue $\isopy =\pm 1$ (red and blue in the figures), analogous to the spin blocks in the Bernevig-Hughes-Zhang model~\cite{Beugeling2021PRB}. In absence of disorder, hybridization can occur only between states in the \emph{same} block.

\textbf{Wave function properties}:
The $\kdotp$ calculations provide access to the wave function probability density $|\psi(y,z)|^2$. By integration over the $z$ coordinate, we find the wave function profile $|\psi(y)|^2=\int |\psi(y,z)|^2 dz$ in the transversal direction. From this wave function profile, we calculate the expectation value $\avg{y}$ and its width $\sigma_y$, defined from the variance $\sigma_y^2=\avg{y^2}-\avg{y}^2$.

For Landau level states, the wave function is approximately Gaussian in shape with mean $\avg{y}=k_x\lB^2$ and standard deviation $\sigma_y = \lB/\sqrt{2}$ (for the lowest Laundau level state), where $\lB=\sqrt{\hbar/eB}$ is the magnetic length. The wave function is deformed near the edge, if the distance between the expected mean and the device edge is less than a few standard deviations,
$w/2 - |k_x|\lB^2\leq c\lB/\sqrt{2}$,
with an arbitrary constant $c\sim 3$ (cf.\ Ref.~\cite{BoettcherEA2020}). For larger $|k_x|$, the wave function is `squeezed' towards the edge which leads to a narrower, more strongly peaked wave function with $\sigma_y<\lB/\sqrt{2}$. The deformation of the wave function also leads to an extra contribution from the Hamiltonian: The energy $E(k_x)$ increases or decreases sharply in this regime.

For $B=5\,\mathrm{T}$, the magnetic length is $\lB=11\nm$, so that $\sigma_y=8\nm$. For a device width of $w=500\nm$, the solution of $|k_x|\lB^2=w/2$ is $|k_x|=1.90\nm^{-1}$. The edge regime, where the mean is less than $c=3$ standard deviations away from the edge, is for $|k_x|\gtrsim 1.71\nm^{-1}$.


\begin{thebibliography}{23}%
\makeatletter
\providecommand \@ifxundefined [1]{%
 \@ifx{#1\undefined}
}%
\providecommand \@ifnum [1]{%
 \ifnum #1\expandafter \@firstoftwo
 \else \expandafter \@secondoftwo
 \fi
}%
\providecommand \@ifx [1]{%
 \ifx #1\expandafter \@firstoftwo
 \else \expandafter \@secondoftwo
 \fi
}%
\providecommand \natexlab [1]{#1}%
\providecommand \enquote  [1]{``#1''}%
\providecommand \bibnamefont  [1]{#1}%
\providecommand \bibfnamefont [1]{#1}%
\providecommand \citenamefont [1]{#1}%
\providecommand \href@noop [0]{\@secondoftwo}%
\providecommand \href [0]{\begingroup \@sanitize@url \@href}%
\providecommand \@href[1]{\@@startlink{#1}\@@href}%
\providecommand \@@href[1]{\endgroup#1\@@endlink}%
\providecommand \@sanitize@url [0]{\catcode `\\12\catcode `\$12\catcode
  `\&12\catcode `\#12\catcode `\^12\catcode `\_12\catcode `\%12\relax}%
\providecommand \@@startlink[1]{}%
\providecommand \@@endlink[0]{}%
\providecommand \url  [0]{\begingroup\@sanitize@url \@url }%
\providecommand \@url [1]{\endgroup\@href {#1}{\urlprefix }}%
\providecommand \urlprefix  [0]{URL }%
\providecommand \Eprint [0]{\href }%
\providecommand \doibase [0]{http://dx.doi.org/}%
\providecommand \selectlanguage [0]{\@gobble}%
\providecommand \bibinfo  [0]{\@secondoftwo}%
\providecommand \bibfield  [0]{\@secondoftwo}%
\providecommand \translation [1]{[#1]}%
\providecommand \BibitemOpen [0]{}%
\providecommand \bibitemStop [0]{}%
\providecommand \bibitemNoStop [0]{.\EOS\space}%
\providecommand \EOS [0]{\spacefactor3000\relax}%
\providecommand \BibitemShut  [1]{\csname bibitem#1\endcsname}%
\let\auto@bib@innerbib\@empty
\bibitem [{\citenamefont {K\"onig}\ \emph {et~al.}(2007)\citenamefont
  {K\"onig}, \citenamefont {Wiedmann}, \citenamefont {Br\"une}, \citenamefont
  {Roth}, \citenamefont {Buhmann}, \citenamefont {Molenkamp}, \citenamefont
  {Qi},\ and\ \citenamefont {Zhang}}]{KonigEA2007}%
  \BibitemOpen
  \bibfield  {author} {\bibinfo {author} {\bibfnamefont {M.}~\bibnamefont
  {K\"onig}}, \bibinfo {author} {\bibfnamefont {S.}~\bibnamefont {Wiedmann}},
  \bibinfo {author} {\bibfnamefont {C.}~\bibnamefont {Br\"une}}, \bibinfo
  {author} {\bibfnamefont {A.}~\bibnamefont {Roth}}, \bibinfo {author}
  {\bibfnamefont {H.}~\bibnamefont {Buhmann}}, \bibinfo {author} {\bibfnamefont
  {L.~W.}\ \bibnamefont {Molenkamp}}, \bibinfo {author} {\bibfnamefont {X.-L.}\
  \bibnamefont {Qi}}, \ and\ \bibinfo {author} {\bibfnamefont {S.-C.}\
  \bibnamefont {Zhang}},\ }\href {\doibase 10.1126/science.1148047} {\bibfield
  {journal} {\bibinfo  {journal} {Science}\ }\textbf {\bibinfo {volume}
  {318}},\ \bibinfo {pages} {766} (\bibinfo {year} {2007})}\BibitemShut
  {NoStop}%
\bibitem [{\citenamefont {Bernevig}\ \emph {et~al.}(2006)\citenamefont
  {Bernevig}, \citenamefont {Hughes},\ and\ \citenamefont
  {Zhang}}]{BernevigEA2006}%
  \BibitemOpen
  \bibfield  {author} {\bibinfo {author} {\bibfnamefont {B.~A.}\ \bibnamefont
  {Bernevig}}, \bibinfo {author} {\bibfnamefont {T.~L.}\ \bibnamefont
  {Hughes}}, \ and\ \bibinfo {author} {\bibfnamefont {S.-C.}\ \bibnamefont
  {Zhang}},\ }\href {\doibase 10.1126/science.1133734} {\bibfield  {journal}
  {\bibinfo  {journal} {Science}\ }\textbf {\bibinfo {volume} {314}},\ \bibinfo
  {pages} {1757} (\bibinfo {year} {2006})}\BibitemShut {NoStop}%
\bibitem [{\citenamefont {Tkachov}\ and\ \citenamefont
  {Hankiewicz}(2010)}]{TkachovHankiewicz2010}%
  \BibitemOpen
  \bibfield  {author} {\bibinfo {author} {\bibfnamefont {G.}~\bibnamefont
  {Tkachov}}\ and\ \bibinfo {author} {\bibfnamefont {E.~M.}\ \bibnamefont
  {Hankiewicz}},\ }\href {\doibase 10.1103/PhysRevLett.104.166803} {\bibfield
  {journal} {\bibinfo  {journal} {Phys. Rev. Lett.}\ }\textbf {\bibinfo
  {volume} {104}},\ \bibinfo {pages} {166803} (\bibinfo {year}
  {2010})}\BibitemShut {NoStop}%
\bibitem [{\citenamefont {Scharf}\ \emph {et~al.}(2012)\citenamefont {Scharf},
  \citenamefont {Matos-Abiague},\ and\ \citenamefont
  {Fabian}}]{ScharfEA2012PRB}%
  \BibitemOpen
  \bibfield  {author} {\bibinfo {author} {\bibfnamefont {B.}~\bibnamefont
  {Scharf}}, \bibinfo {author} {\bibfnamefont {A.}~\bibnamefont
  {Matos-Abiague}}, \ and\ \bibinfo {author} {\bibfnamefont {J.}~\bibnamefont
  {Fabian}},\ }\href {\doibase 10.1103/PhysRevB.86.075418} {\bibfield
  {journal} {\bibinfo  {journal} {Phys. Rev. B}\ }\textbf {\bibinfo {volume}
  {86}},\ \bibinfo {pages} {075418} (\bibinfo {year} {2012})}\BibitemShut
  {NoStop}%
\bibitem [{\citenamefont {Chen}\ \emph {et~al.}(2012)\citenamefont {Chen},
  \citenamefont {Wang},\ and\ \citenamefont {Sun}}]{ChenEA2012PRB}%
  \BibitemOpen
  \bibfield  {author} {\bibinfo {author} {\bibfnamefont {J.-C.}\ \bibnamefont
  {Chen}}, \bibinfo {author} {\bibfnamefont {J.}~\bibnamefont {Wang}}, \ and\
  \bibinfo {author} {\bibfnamefont {Q.-F.}\ \bibnamefont {Sun}},\ }\href
  {\doibase 10.1103/PhysRevB.85.125401} {\bibfield  {journal} {\bibinfo
  {journal} {Phys. Rev. B}\ }\textbf {\bibinfo {volume} {85}},\ \bibinfo
  {pages} {125401} (\bibinfo {year} {2012})}\BibitemShut {NoStop}%
\bibitem [{\citenamefont {V\"ayrynen}\ \emph {et~al.}(2013)\citenamefont
  {V\"ayrynen}, \citenamefont {Goldstein},\ and\ \citenamefont
  {Glazman}}]{VayrynenEA2013}%
  \BibitemOpen
  \bibfield  {author} {\bibinfo {author} {\bibfnamefont {J.~I.}\ \bibnamefont
  {V\"ayrynen}}, \bibinfo {author} {\bibfnamefont {M.}~\bibnamefont
  {Goldstein}}, \ and\ \bibinfo {author} {\bibfnamefont {L.~I.}\ \bibnamefont
  {Glazman}},\ }\href {\doibase 10.1103/PhysRevLett.110.216402} {\bibfield
  {journal} {\bibinfo  {journal} {Phys. Rev. Lett.}\ }\textbf {\bibinfo
  {volume} {110}},\ \bibinfo {pages} {216402} (\bibinfo {year}
  {2013})}\BibitemShut {NoStop}%
\bibitem [{\citenamefont {V\"ayrynen}\ \emph {et~al.}(2014)\citenamefont
  {V\"ayrynen}, \citenamefont {Goldstein}, \citenamefont {Gefen},\ and\
  \citenamefont {Glazman}}]{VayrynenEA2014}%
  \BibitemOpen
  \bibfield  {author} {\bibinfo {author} {\bibfnamefont {J.~I.}\ \bibnamefont
  {V\"ayrynen}}, \bibinfo {author} {\bibfnamefont {M.}~\bibnamefont
  {Goldstein}}, \bibinfo {author} {\bibfnamefont {Y.}~\bibnamefont {Gefen}}, \
  and\ \bibinfo {author} {\bibfnamefont {L.~I.}\ \bibnamefont {Glazman}},\
  }\href {\doibase 10.1103/PhysRevB.90.115309} {\bibfield  {journal} {\bibinfo
  {journal} {Phys. Rev. B}\ }\textbf {\bibinfo {volume} {90}},\ \bibinfo
  {pages} {115309} (\bibinfo {year} {2014})}\BibitemShut {NoStop}%
\bibitem [{\citenamefont {Ma}\ \emph {et~al.}(2015)\citenamefont {Ma},
  \citenamefont {Calvo}, \citenamefont {Wang}, \citenamefont {Lian},
  \citenamefont {M\"{u}hlbauer}, \citenamefont {Br\"{u}ne}, \citenamefont
  {Cui}, \citenamefont {Lai}, \citenamefont {Kundhikanjana}, \citenamefont
  {Yang}, \citenamefont {Baenninger}, \citenamefont {K\"{o}nig}, \citenamefont
  {Ames}, \citenamefont {Buhmann}, \citenamefont {Leubner}, \citenamefont
  {Molenkamp}, \citenamefont {Zhang}, \citenamefont {Goldhaber-Gordon},
  \citenamefont {Kelly},\ and\ \citenamefont {Shen}}]{MaEA2015NatComm}%
  \BibitemOpen
  \bibfield  {author} {\bibinfo {author} {\bibfnamefont {E.~Y.}\ \bibnamefont
  {Ma}}, \bibinfo {author} {\bibfnamefont {M.~R.}\ \bibnamefont {Calvo}},
  \bibinfo {author} {\bibfnamefont {J.}~\bibnamefont {Wang}}, \bibinfo {author}
  {\bibfnamefont {B.}~\bibnamefont {Lian}}, \bibinfo {author} {\bibfnamefont
  {M.}~\bibnamefont {M\"{u}hlbauer}}, \bibinfo {author} {\bibfnamefont
  {C.}~\bibnamefont {Br\"{u}ne}}, \bibinfo {author} {\bibfnamefont {Y.-T.}\
  \bibnamefont {Cui}}, \bibinfo {author} {\bibfnamefont {K.}~\bibnamefont
  {Lai}}, \bibinfo {author} {\bibfnamefont {W.}~\bibnamefont {Kundhikanjana}},
  \bibinfo {author} {\bibfnamefont {Y.}~\bibnamefont {Yang}}, \bibinfo {author}
  {\bibfnamefont {M.}~\bibnamefont {Baenninger}}, \bibinfo {author}
  {\bibfnamefont {M.}~\bibnamefont {K\"{o}nig}}, \bibinfo {author}
  {\bibfnamefont {C.}~\bibnamefont {Ames}}, \bibinfo {author} {\bibfnamefont
  {H.}~\bibnamefont {Buhmann}}, \bibinfo {author} {\bibfnamefont
  {P.}~\bibnamefont {Leubner}}, \bibinfo {author} {\bibfnamefont {L.~W.}\
  \bibnamefont {Molenkamp}}, \bibinfo {author} {\bibfnamefont {S.-C.}\
  \bibnamefont {Zhang}}, \bibinfo {author} {\bibfnamefont {D.}~\bibnamefont
  {Goldhaber-Gordon}}, \bibinfo {author} {\bibfnamefont {M.~A.}\ \bibnamefont
  {Kelly}}, \ and\ \bibinfo {author} {\bibfnamefont {Z.-X.}\ \bibnamefont
  {Shen}},\ }\href {\doibase 10.1038/ncomms8252} {\bibfield  {journal}
  {\bibinfo  {journal} {Nat. Commun.}\ }\textbf {\bibinfo {volume} {6}},\
  \bibinfo {pages} {7252} (\bibinfo {year} {2015})}\BibitemShut {NoStop}%
\bibitem [{\citenamefont {Bendias}\ \emph {et~al.}(2018)\citenamefont
  {Bendias}, \citenamefont {Shamim}, \citenamefont {Herrmann}, \citenamefont
  {Budewitz}, \citenamefont {Shekhar}, \citenamefont {Leubner}, \citenamefont
  {Kleinlein}, \citenamefont {Bocquillon}, \citenamefont {Buhmann},\ and\
  \citenamefont {Molenkamp}}]{BendiasEA2018}%
  \BibitemOpen
  \bibfield  {author} {\bibinfo {author} {\bibfnamefont {K.}~\bibnamefont
  {Bendias}}, \bibinfo {author} {\bibfnamefont {S.}~\bibnamefont {Shamim}},
  \bibinfo {author} {\bibfnamefont {O.}~\bibnamefont {Herrmann}}, \bibinfo
  {author} {\bibfnamefont {A.}~\bibnamefont {Budewitz}}, \bibinfo {author}
  {\bibfnamefont {P.}~\bibnamefont {Shekhar}}, \bibinfo {author} {\bibfnamefont
  {P.}~\bibnamefont {Leubner}}, \bibinfo {author} {\bibfnamefont
  {J.}~\bibnamefont {Kleinlein}}, \bibinfo {author} {\bibfnamefont
  {E.}~\bibnamefont {Bocquillon}}, \bibinfo {author} {\bibfnamefont
  {H.}~\bibnamefont {Buhmann}}, \ and\ \bibinfo {author} {\bibfnamefont
  {L.~W.}\ \bibnamefont {Molenkamp}},\ }\href {\doibase
  10.1021/acs.nanolett.8b01405} {\bibfield  {journal} {\bibinfo  {journal}
  {Nano Letters}\ }\textbf {\bibinfo {volume} {18}},\ \bibinfo {pages} {4831}
  (\bibinfo {year} {2018})}\BibitemShut {NoStop}%
\bibitem [{\citenamefont {Lafont}\ \emph {et~al.}(2019)\citenamefont {Lafont},
  \citenamefont {Rosenblatt}, \citenamefont {Heiblum},\ and\ \citenamefont
  {Umansky}}]{LafontEA2019Science}%
  \BibitemOpen
  \bibfield  {author} {\bibinfo {author} {\bibfnamefont {F.}~\bibnamefont
  {Lafont}}, \bibinfo {author} {\bibfnamefont {A.}~\bibnamefont {Rosenblatt}},
  \bibinfo {author} {\bibfnamefont {M.}~\bibnamefont {Heiblum}}, \ and\
  \bibinfo {author} {\bibfnamefont {V.}~\bibnamefont {Umansky}},\ }\href
  {\doibase 10.1126/science.aar3766} {\bibfield  {journal} {\bibinfo  {journal}
  {Science}\ }\textbf {\bibinfo {volume} {363}},\ \bibinfo {pages} {54}
  (\bibinfo {year} {2019})}\BibitemShut {NoStop}%
\bibitem [{\citenamefont {MacDonald}(1990)}]{MacDonald1990PRL}%
  \BibitemOpen
  \bibfield  {author} {\bibinfo {author} {\bibfnamefont {A.~H.}\ \bibnamefont
  {MacDonald}},\ }\href {\doibase 10.1103/PhysRevLett.64.220} {\bibfield
  {journal} {\bibinfo  {journal} {Phys. Rev. Lett.}\ }\textbf {\bibinfo
  {volume} {64}},\ \bibinfo {pages} {220} (\bibinfo {year} {1990})}\BibitemShut
  {NoStop}%
\bibitem [{\citenamefont {Moore}\ and\ \citenamefont
  {Haldane}(1997)}]{Moore_Haldane_1997PRB}%
  \BibitemOpen
  \bibfield  {author} {\bibinfo {author} {\bibfnamefont {J.~E.}\ \bibnamefont
  {Moore}}\ and\ \bibinfo {author} {\bibfnamefont {F.~D.~M.}\ \bibnamefont
  {Haldane}},\ }\href {\doibase 10.1103/PhysRevB.55.7818} {\bibfield  {journal}
  {\bibinfo  {journal} {Phys. Rev. B}\ }\textbf {\bibinfo {volume} {55}},\
  \bibinfo {pages} {7818} (\bibinfo {year} {1997})}\BibitemShut {NoStop}%
\bibitem [{\citenamefont {Shamim}\ \emph {et~al.}(2020)\citenamefont {Shamim},
  \citenamefont {Beugeling}, \citenamefont {B{\"o}ttcher}, \citenamefont
  {Shekhar}, \citenamefont {Budewitz}, \citenamefont {Leubner}, \citenamefont
  {Lunczer}, \citenamefont {Hankiewicz}, \citenamefont {Buhmann},\ and\
  \citenamefont {Molenkamp}}]{ShamimEA2020SciAdv}%
  \BibitemOpen
  \bibfield  {author} {\bibinfo {author} {\bibfnamefont {S.}~\bibnamefont
  {Shamim}}, \bibinfo {author} {\bibfnamefont {W.}~\bibnamefont {Beugeling}},
  \bibinfo {author} {\bibfnamefont {J.}~\bibnamefont {B{\"o}ttcher}}, \bibinfo
  {author} {\bibfnamefont {P.}~\bibnamefont {Shekhar}}, \bibinfo {author}
  {\bibfnamefont {A.}~\bibnamefont {Budewitz}}, \bibinfo {author}
  {\bibfnamefont {P.}~\bibnamefont {Leubner}}, \bibinfo {author} {\bibfnamefont
  {L.}~\bibnamefont {Lunczer}}, \bibinfo {author} {\bibfnamefont {E.~M.}\
  \bibnamefont {Hankiewicz}}, \bibinfo {author} {\bibfnamefont
  {H.}~\bibnamefont {Buhmann}}, \ and\ \bibinfo {author} {\bibfnamefont
  {L.~W.}\ \bibnamefont {Molenkamp}},\ }\href {\doibase 10.1126/sciadv.aba4625}
  {\bibfield  {journal} {\bibinfo  {journal} {Sci. Adv.}\ }\textbf {\bibinfo
  {volume} {6}} (\bibinfo {year} {2020}),\ 10.1126/sciadv.aba4625}\BibitemShut
  {NoStop}%
\bibitem [{\citenamefont {Beugeling}(2021)}]{Beugeling2021PRB}%
  \BibitemOpen
  \bibfield  {author} {\bibinfo {author} {\bibfnamefont {W.}~\bibnamefont
  {Beugeling}},\ }\href {\doibase 10.1103/PhysRevB.104.115428} {\bibfield
  {journal} {\bibinfo  {journal} {Phys. Rev. B}\ }\textbf {\bibinfo {volume}
  {104}},\ \bibinfo {pages} {115428} (\bibinfo {year} {2021})}\BibitemShut
  {NoStop}%
\bibitem [{\citenamefont {Shamim}\ \emph {et~al.}(2021)\citenamefont {Shamim},
  \citenamefont {Beugeling}, \citenamefont {Shekhar}, \citenamefont {Bendias},
  \citenamefont {Lunczer}, \citenamefont {Kleinlein}, \citenamefont {Buhmann},\
  and\ \citenamefont {Molenkamp}}]{ShamimEA2021NatComm}%
  \BibitemOpen
  \bibfield  {author} {\bibinfo {author} {\bibfnamefont {S.}~\bibnamefont
  {Shamim}}, \bibinfo {author} {\bibfnamefont {W.}~\bibnamefont {Beugeling}},
  \bibinfo {author} {\bibfnamefont {P.}~\bibnamefont {Shekhar}}, \bibinfo
  {author} {\bibfnamefont {K.}~\bibnamefont {Bendias}}, \bibinfo {author}
  {\bibfnamefont {L.}~\bibnamefont {Lunczer}}, \bibinfo {author} {\bibfnamefont
  {J.}~\bibnamefont {Kleinlein}}, \bibinfo {author} {\bibfnamefont
  {H.}~\bibnamefont {Buhmann}}, \ and\ \bibinfo {author} {\bibfnamefont
  {L.~W.}\ \bibnamefont {Molenkamp}},\ }\href {\doibase
  10.1038/s41467-021-23262-1} {\bibfield  {journal} {\bibinfo  {journal} {Nat.
  Commun.}\ }\textbf {\bibinfo {volume} {12}},\ \bibinfo {pages} {3193}
  (\bibinfo {year} {2021})}\BibitemShut {NoStop}%
\bibitem [{\citenamefont {Roth}\ \emph {et~al.}(2009)\citenamefont {Roth},
  \citenamefont {Br\"une}, \citenamefont {Buhmann}, \citenamefont {Molenkamp},
  \citenamefont {Maciejko}, \citenamefont {Qi},\ and\ \citenamefont
  {Zhang}}]{RothEA2009}%
  \BibitemOpen
  \bibfield  {author} {\bibinfo {author} {\bibfnamefont {A.}~\bibnamefont
  {Roth}}, \bibinfo {author} {\bibfnamefont {C.}~\bibnamefont {Br\"une}},
  \bibinfo {author} {\bibfnamefont {H.}~\bibnamefont {Buhmann}}, \bibinfo
  {author} {\bibfnamefont {L.~W.}\ \bibnamefont {Molenkamp}}, \bibinfo {author}
  {\bibfnamefont {J.}~\bibnamefont {Maciejko}}, \bibinfo {author}
  {\bibfnamefont {X.-L.}\ \bibnamefont {Qi}}, \ and\ \bibinfo {author}
  {\bibfnamefont {S.-C.}\ \bibnamefont {Zhang}},\ }\href {\doibase
  10.1126/science.1174736} {\bibfield  {journal} {\bibinfo  {journal}
  {Science}\ }\textbf {\bibinfo {volume} {325}},\ \bibinfo {pages} {294}
  (\bibinfo {year} {2009})}\BibitemShut {NoStop}%
\bibitem [{\citenamefont {B\"uttiker}(1986)}]{Buttiker1986PRL}%
  \BibitemOpen
  \bibfield  {author} {\bibinfo {author} {\bibfnamefont {M.}~\bibnamefont
  {B\"uttiker}},\ }\href {\doibase 10.1103/PhysRevLett.57.1761} {\bibfield
  {journal} {\bibinfo  {journal} {Phys. Rev. Lett.}\ }\textbf {\bibinfo
  {volume} {57}},\ \bibinfo {pages} {1761} (\bibinfo {year}
  {1986})}\BibitemShut {NoStop}%
\bibitem [{\citenamefont {Bernevig}\ and\ \citenamefont
  {Zhang}(2006)}]{BernevigZhang2006}%
  \BibitemOpen
  \bibfield  {author} {\bibinfo {author} {\bibfnamefont {B.~A.}\ \bibnamefont
  {Bernevig}}\ and\ \bibinfo {author} {\bibfnamefont {S.-C.}\ \bibnamefont
  {Zhang}},\ }\href {\doibase 10.1103/PhysRevLett.96.106802} {\bibfield
  {journal} {\bibinfo  {journal} {Phys. Rev. Lett.}\ }\textbf {\bibinfo
  {volume} {96}},\ \bibinfo {pages} {106802} (\bibinfo {year}
  {2006})}\BibitemShut {NoStop}%
\bibitem [{\citenamefont {B\"ottcher}\ \emph {et~al.}(2019)\citenamefont
  {B\"ottcher}, \citenamefont {Tutschku}, \citenamefont {Molenkamp},\ and\
  \citenamefont {Hankiewicz}}]{BoettcherEA2019}%
  \BibitemOpen
  \bibfield  {author} {\bibinfo {author} {\bibfnamefont {J.}~\bibnamefont
  {B\"ottcher}}, \bibinfo {author} {\bibfnamefont {C.}~\bibnamefont
  {Tutschku}}, \bibinfo {author} {\bibfnamefont {L.~W.}\ \bibnamefont
  {Molenkamp}}, \ and\ \bibinfo {author} {\bibfnamefont {E.~M.}\ \bibnamefont
  {Hankiewicz}},\ }\href {\doibase 10.1103/PhysRevLett.123.226602} {\bibfield
  {journal} {\bibinfo  {journal} {Phys. Rev. Lett.}\ }\textbf {\bibinfo
  {volume} {123}},\ \bibinfo {pages} {226602} (\bibinfo {year}
  {2019})}\BibitemShut {NoStop}%
\bibitem [{\citenamefont {B\"ottcher}\ \emph {et~al.}(2020)\citenamefont
  {B\"ottcher}, \citenamefont {Tutschku},\ and\ \citenamefont
  {Hankiewicz}}]{BoettcherEA2020}%
  \BibitemOpen
  \bibfield  {author} {\bibinfo {author} {\bibfnamefont {J.}~\bibnamefont
  {B\"ottcher}}, \bibinfo {author} {\bibfnamefont {C.}~\bibnamefont
  {Tutschku}}, \ and\ \bibinfo {author} {\bibfnamefont {E.~M.}\ \bibnamefont
  {Hankiewicz}},\ }\href {\doibase 10.1103/PhysRevB.101.195433} {\bibfield
  {journal} {\bibinfo  {journal} {Phys. Rev. B}\ }\textbf {\bibinfo {volume}
  {101}},\ \bibinfo {pages} {195433} (\bibinfo {year} {2020})}\BibitemShut
  {NoStop}%
\bibitem [{\citenamefont {Lunczer}\ \emph {et~al.}(2019)\citenamefont
  {Lunczer}, \citenamefont {Leubner}, \citenamefont {Endres}, \citenamefont
  {M\"uller}, \citenamefont {Br\"une}, \citenamefont {Buhmann},\ and\
  \citenamefont {Molenkamp}}]{LunczerEA2019PRL}%
  \BibitemOpen
  \bibfield  {author} {\bibinfo {author} {\bibfnamefont {L.}~\bibnamefont
  {Lunczer}}, \bibinfo {author} {\bibfnamefont {P.}~\bibnamefont {Leubner}},
  \bibinfo {author} {\bibfnamefont {M.}~\bibnamefont {Endres}}, \bibinfo
  {author} {\bibfnamefont {V.~L.}\ \bibnamefont {M\"uller}}, \bibinfo {author}
  {\bibfnamefont {C.}~\bibnamefont {Br\"une}}, \bibinfo {author} {\bibfnamefont
  {H.}~\bibnamefont {Buhmann}}, \ and\ \bibinfo {author} {\bibfnamefont
  {L.~W.}\ \bibnamefont {Molenkamp}},\ }\href {\doibase
  10.1103/PhysRevLett.123.047701} {\bibfield  {journal} {\bibinfo  {journal}
  {Phys. Rev. Lett.}\ }\textbf {\bibinfo {volume} {123}},\ \bibinfo {pages}
  {047701} (\bibinfo {year} {2019})}\BibitemShut {NoStop}%
\bibitem [{\citenamefont {K\"onig}\ \emph {et~al.}(2013)\citenamefont
  {K\"onig}, \citenamefont {Baenninger}, \citenamefont {Garcia}, \citenamefont
  {Harjee}, \citenamefont {Pruitt}, \citenamefont {Ames}, \citenamefont
  {Leubner}, \citenamefont {Br\"une}, \citenamefont {Buhmann}, \citenamefont
  {Molenkamp},\ and\ \citenamefont {Goldhaber-Gordon}}]{KonigEA2013PRX}%
  \BibitemOpen
  \bibfield  {author} {\bibinfo {author} {\bibfnamefont {M.}~\bibnamefont
  {K\"onig}}, \bibinfo {author} {\bibfnamefont {M.}~\bibnamefont {Baenninger}},
  \bibinfo {author} {\bibfnamefont {A.~G.~F.}\ \bibnamefont {Garcia}}, \bibinfo
  {author} {\bibfnamefont {N.}~\bibnamefont {Harjee}}, \bibinfo {author}
  {\bibfnamefont {B.~L.}\ \bibnamefont {Pruitt}}, \bibinfo {author}
  {\bibfnamefont {C.}~\bibnamefont {Ames}}, \bibinfo {author} {\bibfnamefont
  {P.}~\bibnamefont {Leubner}}, \bibinfo {author} {\bibfnamefont
  {C.}~\bibnamefont {Br\"une}}, \bibinfo {author} {\bibfnamefont
  {H.}~\bibnamefont {Buhmann}}, \bibinfo {author} {\bibfnamefont {L.~W.}\
  \bibnamefont {Molenkamp}}, \ and\ \bibinfo {author} {\bibfnamefont
  {D.}~\bibnamefont {Goldhaber-Gordon}},\ }\href {\doibase
  10.1103/PhysRevX.3.021003} {\bibfield  {journal} {\bibinfo  {journal} {Phys.
  Rev. X}\ }\textbf {\bibinfo {volume} {3}},\ \bibinfo {pages} {021003}
  (\bibinfo {year} {2013})}\BibitemShut {NoStop}%
\bibitem [{\citenamefont {Kane}(1957)}]{Kane1957}%
  \BibitemOpen
  \bibfield  {author} {\bibinfo {author} {\bibfnamefont {E.~O.}\ \bibnamefont
  {Kane}},\ }\href {\doibase 10.1016/0022-3697(57)90013-6} {\bibfield
  {journal} {\bibinfo  {journal} {J. Phys. Chem. Solids}\ }\textbf {\bibinfo
  {volume} {1}},\ \bibinfo {pages} {249} (\bibinfo {year} {1957})}\BibitemShut
  {NoStop}%
\end{thebibliography}


%

\subsection*{Acknowledgements}

We thank C. Gould, C. Morais Smith and C. Br\"{u}ne for useful discussions. We acknowledge financial support from the Deutsche Forschungsgemeinschaft (DFG, German Research Foundation) in the Leibniz Program and in the projects SFB 1170 (Project ID 258499086) and SPP 1666 (Project ID 220179758), from the EU ERC-AdG program (Project 4-TOPS), from the W\"urzburg-Dresden Cluster of Excellence on Complexity and Topology
in Quantum Matter (EXC 2147, Project ID 39085490), and from the Free State of Bavaria (Elitenetzwerk Bayern IDK `Topologische Isolatoren' and the Institute for Topological Insulators).

\subsection*{Author contributions}

S.S., H.B., and L.W.M. planned the experiments. S.S., P.S., and A.B. conducted the measurements and analyzed the data.  The band structure analysis was performed by W.B., J.B., and J.B.M. The material was grown by L.L. The devices were fabricated by P.S. All authors contributed to interpretation of the results. L.W.M., H.B., and E.M.H. supervised the project. S.S. and W.B. wrote the article with input from all authors. S.S., P.S., and W.B. contributed euqally to this work.

\subsection*{Competing interests}

The authors declare no competing interests.

\subsection*{Data availability}

All data necessary to support the conclusions of the paper are available in the manuscript. Additional data is available from the corresponding authors upon reasonable request.

\subsection*{Code availability}

The code used for the band structure analysis is available from the corresponding authors upon reasonable request.

\clearpage

\onecolumngrid
\newpage
\begin{center}
{\noindent\large\bf Supplementary Materials for\\  \emph{Counterpropagating topological and quantum Hall edge channels}}
\vspace{1em}
\end{center}

\setcounter{page}{1} \renewcommand{\thepage}{S\arabic{page}}

\setcounter{figure}{0}   \renewcommand{\thefigure}{S\arabic{figure}}

\setcounter{equation}{0} \renewcommand{\theequation}{S.\arabic{equation}}

\setcounter{table}{0} \renewcommand{\thetable}{S.\arabic{table}}

\setcounter{section}{0} \renewcommand{\thesection}{S\arabic{section}}

\renewcommand{\thesubsection}{S\arabic{section}.\Alph{subsection}}


\makeatletter
\renewcommand*{\p@subsection}{}
\makeatother

\renewcommand{\thesubsubsection}{S\arabic{section}.\Alph{subsection}-\arabic{subsubsection}}

\makeatletter
\renewcommand*{\p@subsubsection}{}  
\makeatother



\textbf{\large Supplementary Note 1: The quantum spin Hall effect in (Hg,Mn)Te quantum wells}
\vspace{0.5cm}

Supplementary Figure 1a and c show the gate voltage and magnetotransport characteristics of microstructures fabricated from QW$4$ at $1.4$\,K. Supplementary Figure 1a shows that we can tune the chemical potential from $n$- to $p$-conduction regime by decreasing the gate voltage $V_g$. For $V_g$ in the range of $-0.1$ to $-0.25$\,V, the conductance is quantized close to $2e^2/h$ (dashed line in Supplementary Figure 1a) due to the quantum spin Hall effect. Similar conductance quantization due to the quantum spin Hall effect is also seen in microstructures fabricated from QW$5$ (Supplementary Figure 1b). The inset of Supplementary Figure 1c shows the conductance quantization for Dev 4.w$1$ at $20$\,mK (this result is already published in Ref.~14 of the main text). The long quantized plateau in conductance observed in Supplementary Figure 1b and inset of Supplementary Figure 1c is due to the pinning of the chemical potential to the `camel back' feature in the valence band (due to band inversion-induced van Hove singularity in the valence band of topological (Hg,Mn)Te quantum wells), as already explained in Ref.~14 of the main text. For Dev 4.w$1$, at $V_g=-0.2$\,V (where the conductance is quantized to $2e^2/h$ at $B=0$), application of a perpendicular magnetic field ($> 160$\,mT) results in the quantization of the transverse resistance $R_{xy}$ to $-h/e^2$ (solid line in Supplementary Figure 1c) and simultaneously $R_{xx}$ goes to zero. This low field $\nu=-1$ plateau is the emergent quantum Hall plateau, which emerges from the quantum spin Hall edge channels, as discussed previously in Ref.~13 of the main text.

\clearpage

\textbf{\large Supplementary Figures}

\vspace{0.5cm}


\begin{figure*}[!ht]
\includegraphics[width=0.85\linewidth]{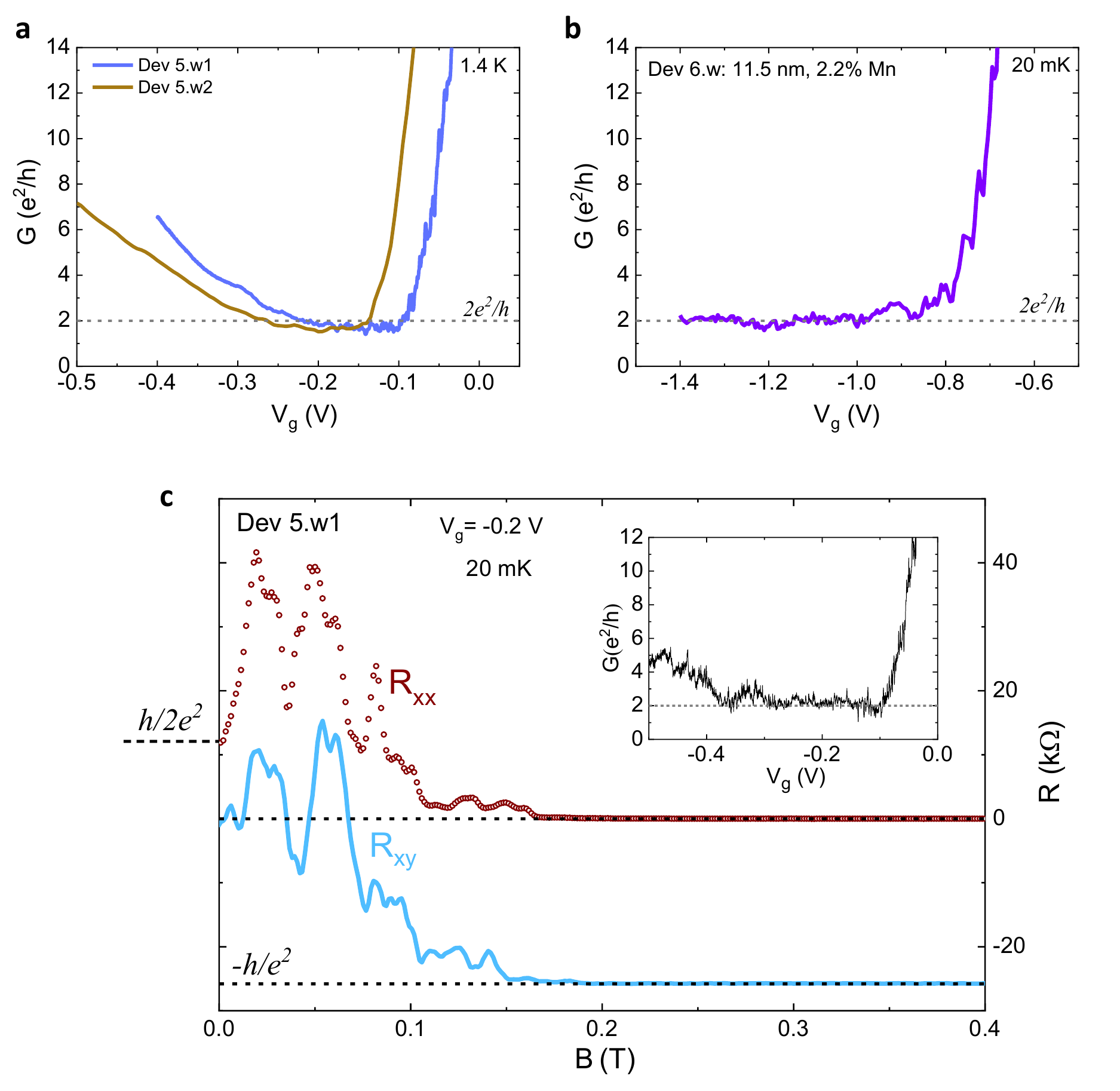}
\end{figure*}
{\noindent \textbf{Supplementary Figure 1: Quantized conductance in (Hg,Mn)Te quantum wells and transition to the emergent quantum Hall effect.}
The conductance $G$, as a function of gate voltage $V_g$ for
\textbf{a,} Dev 4.w$1$ and Dev 4.w$2$ at $1.4$\,K.
\textbf{b,} Dev 5.w at $20$\,mK.
\textbf{c,} Longitudinal (open circles) and transverse resistance (solid line), $R_{xx}$ and $R_{xy}$ respectively of Dev 4.w$1$, as a function of magnetic field $B$ at $20$\,mK for $V_g=-0.2$\,V. The inset shows $G$ as a function of $V_g$.}


\clearpage


\begin{figure*}[!ht]
\includegraphics[width=1\linewidth]{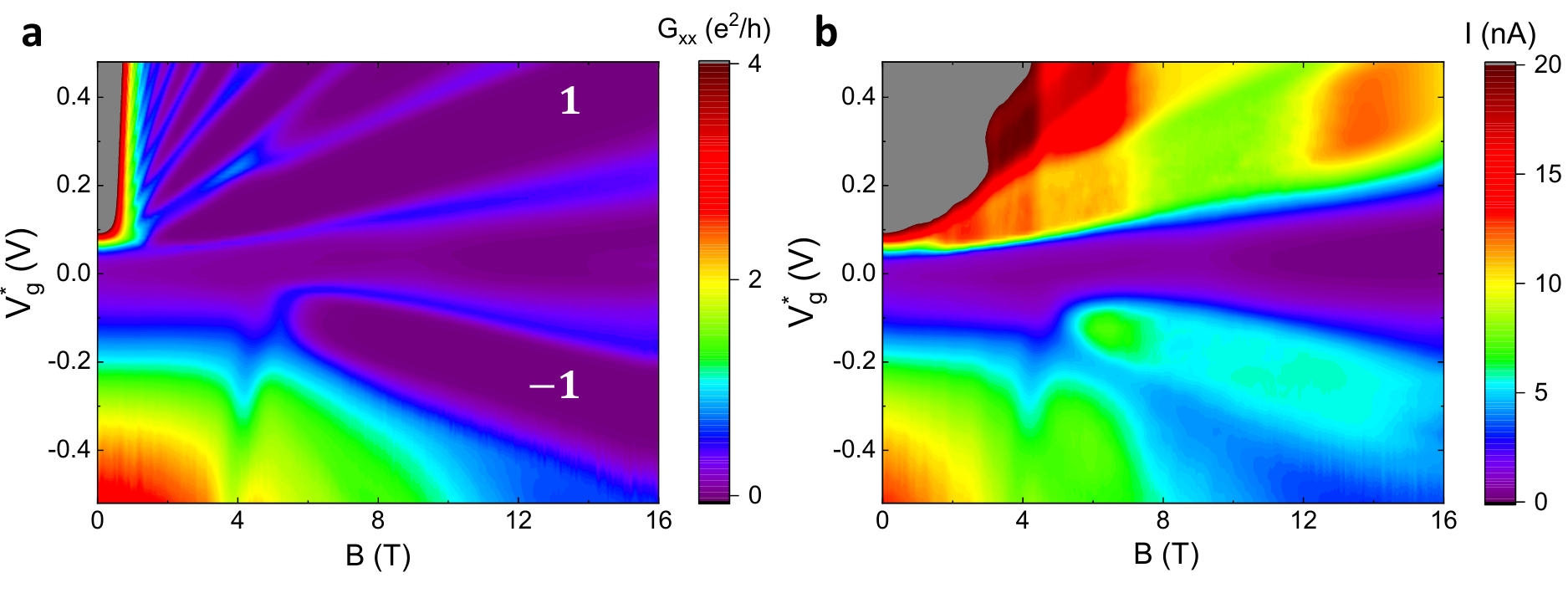}
\end{figure*}
{\noindent \textbf{Supplementary Figure 2: Magnetotransport for wet-etched device Dev 2.w.}
\textbf{a,} Color plot of the longitudinal conductance $G_{xx}$ as a function of $B$ and $V_g^*$.
\textbf{b,} Color plot of the current $I$ flowing through the device as a function of $B$ and $V_g^*$.}


\vspace{1cm}

\end{document}